# Static magnetic field concentration and enhancement using magnetic materials with positive permeability

F. Sun[1, 2] and S. He[1, 2*]

1. Department of Electromagnetic Engineering, School of Electrical Engineering, Royal Institute of Technology (KTH)
2. Centre for Optical and Electromagnetic Research, JORCEP

**Abstract**
   In this paper a novel compressor for static magnetic fields is proposed based on finite embedded transformation optics. When the DC magnetic field passes through the designed device, the magnetic field can be compressed inside the device. After it passes through the device, one can obtain an enhanced static magnetic field behind the output surface of the device (in a free space region). We can also combine our compressor with some other structures to get a higher static magnetic field enhancement in a free space region. In contrast with other devices based on transformation optics for enhancing static magnetic fields, our device is not a closed structure and thus has some special applications (e.g., for controlling magnetic nano-particles for gene and drag delivery). The designed compressor can be constructed by using currently available materials or DC meta-materials with positive permeability. Numerical simulation verifies good performance of our device.

## 1. Introduction

   The static magnetic fields play an essential role in many applications, including magnetic resonance imaging [1], transcranial magnetic stimulation [2], controlling magnetic nano-particles for gene and drag delivery [3, 4], magnetic sensors [5] and so on. Higher static magnetic fields can promote the development of these applications. We can classify the static magnetic field enhancement under two types: one is for enhancing static magnetic fields in a large free space region which may be applied for e.g. improving the spatial resolution in MRI; the other is for enhancing static magnetic fields in a small region which can be used to improve the sensitivity of magnetic sensors and promote the development of magnetic nano-particles for gene and drag delivery, etc.

   Magnetic lenses are passive devices which can cause an external magnetic field to converge/focus without consuming additional power. We can acquire an enhanced DC magnetic field without consuming power by using magnetic lenses, and this is different from magnets which need a power supply to acquire a higher DC magnetic field. A compact high field system (>10T) is an important trend in magnetic enhancement [6, 7]. Based on the diamagnetism of the superconductor, people have designed various magnetic lenses composed of superconductors to concentrate the DC magnetic field when applying a background magnetic field to the lenses [6-10].

However, magnetic lenses based on superconductors have many disadvantages: the degree of enhancement is very low (e.g., magnification factor is only 1.82 by using NbTi/Nb/Cu superconducting multilayered sheets [8], and only 3.2 at the center by using HTS bulk cylinders [9]), refrigeration is required and the performance is often influenced by quenching and other factors.

Transformation optics (TO) is a powerful tool, which is based on the invariance form of Maxwell's equations [10, 11]. A special medium (often referred to as the transformed medium), which can be used to control the trace of the electromagnetic wave and also the static electric or magnetic field, can be designed by TO. As the transformed medium is often complex, which may often not exist in nature, people have to design some artificial materials (e.g., meta-materials) to create them. In the development of DC magnetic and electric meta-materials [12, 13], many DC magnetic and electric devices based on TO (e.g., DC magnetic or electric cloaking devices [13, 14] and DC magnetic or electric energy concentrators [15-17]) have been studied in recent years. In this paper, we design a novel magnetic compressor which can compress the DC magnetic field and obtain a high magnetic field in a free space region. Our device can be constructed with natural materials or current DC magnetic meta-materials [12, 18] with permeability $\mu>0$. Note that it is not clear yet to this community how to achieve a meta-material with $\mu<0$ at DC and fortunately permeability $\mu$ is always positive in our design. We will thoroughly describe the design method in section 2. We will also give numerical simulation results based on the finite element method (FEM), which shows good device performance. In section 3, we combine our magnetic compressor with some other structures to get a further degree of static magnetic field enhancement in a free space. The summary will be given in section 4, where we also make a comparison between our compressor and other devices which can also be used for enhancing the DC magnetic field.

## 2. The method of designing a compressor for static magnetic fields

Traditional transformations are often continuous, and thus the electromagnetic properties of the incident waves are altered exclusively within a restricted region and the corresponding devices are inherently invisible to observers outside of the transformed region (e.g., invisible cloaks and field rotators). Finite embedded transformation (FET) is a special kind of transformations, which is discontinuous at the output surface of the devices [19]. FET can transfer/manipulate fields from the transformed medium to an un-transformed medium, which has been widely used to design light splitters [20], polarization controllers [21], light beam expanders or compressors [22], etc. In this paper, we will use FET to design a magnetic compressor for the static magnetic field, which may have important applications in DC magnetic flux controlling and DC magnetic field enhancement.

For simplicity we consider a two-dimensional (2D) device; the input surface of the device is $x'=0$ and the output surface of the device is $x'=d$. We use ($x'$, $y'$, $z'$) and ($x$, $y$, $z$) to express the coordinate systems in the physical space and reference space, respectively. We can use the following transformation to concentrate magnetic flux at the output of the device:

$$x' = x$$
$$y' = \left[\frac{(M-1)}{d}x + 1\right]y := \alpha y \qquad (1)$$
$$z' = z$$

*where d is the thickness of the device and M (0<M<1) determines both the degree of the enhancement and the permeability distribution of the device. The smaller the value of M, the higher the concentration and also the larger the field enhancement.* The Jacobean matrix of the above transformation can be calculated by

$$\overline{\overline{A}} = \begin{bmatrix} 1 & 0 & 0 \\ \frac{(M-1)}{d}y & \alpha & 0 \\ 0 & 0 & 1 \end{bmatrix} \qquad (2)$$

As the transformation optics is based on the invariant form of Maxwell's equations, it still holds for the static magnetic field cases for which only two of the Maxwell's equations are involved [17]. We also assume that the reference space is free space. The magnetic permeability parameters corresponding to the above transformation can be given:

$$\overline{\overline{\mu}} = \mu_0 \overline{\overline{A}}\,\overline{\overline{A}}^T / \det(\overline{\overline{A}}) = \mu_0 \begin{bmatrix} 1/\alpha & \frac{(M-1)}{d\alpha}y & 0 \\ \frac{(M-1)}{d\alpha}y & \frac{(M-1)^2}{d^2\alpha}y^2 + \alpha & 0 \\ 0 & 0 & 1/\alpha \end{bmatrix} \qquad (3)$$

For a 2D device, only $\mu_{xx}, \mu_{xy}$ and $\mu_{yy}$ affect the magnetic field. We can rewrite Eq. (3) as:

$$\overline{\overline{\mu}} = \mu_0 \begin{bmatrix} 1/\alpha & \frac{(M-1)}{d\alpha}y \\ \frac{(M-1)}{d\alpha}y & \frac{(M-1)^2}{d^2\alpha}y^2 + \alpha \end{bmatrix} := \begin{bmatrix} \mu_{xx} & \mu_{xy} \\ \mu_{xy} & \mu_{yy} \end{bmatrix} \qquad (4)$$

For fabrication convenience of the compressor, it is preferable to denote the material parameters in their eigenbasis, where the permeability tensor is diagonal. Due to the symmetry of the transformed medium, this can always be accomplished. We can diagonalize the permeability tensor in Eq. (4) as:

$$\overline{\overline{\mu}} = diag(\frac{\mu_{xx} + \mu_{yy}}{2} - \kappa, \frac{\mu_{xx} + \mu_{yy}}{2} + \kappa) := diag(\mu_1, \mu_2) \qquad (5)$$

where $\kappa = \dfrac{\sqrt{\mu_{xx}^2 + \mu_{xy}^2 - 2\mu_{xx}\mu_{yy} + 4\mu_{xy}^2}}{2}$

Next we use the finite element method (FEM) [23] to simulate the performance of the proposed device. As shown in Fig. 1(a), 1(d) and 1(g), when a uniformed background static magnetic field $B_b$=1T is imposed onto the device from left to right, the magnetic field is compressed inside the device. At the output surface of the device, we get a concentrated DC magnetic field. We also verify that *M* determines the rate of concentration: In Fig. 1(a), *M*=0.1 and the enhancement factor is about 4.7; In Fig. 1(d), *M*=0.35 and the enhancement factor is about 1.65; In Fig. 1(g), *M*=0.01 and the enhancement factor is about 45. The corresponding permeability of these devices are also shown in Fig. 1(b) and1(c) for *M*=0.1, Fig. 1(e) and 1(f) for *M*=0.35, and Fig. 1(h) and 1(i) for *M*=0.01. Here the permeability is always positive, which can be constructed with natural materials (e.g., diamagnetic substances and paramagnetic materials) or current meta-materials for static magnetic fields [12, 18]. As *M* approaches zero, the concentration degree and enhancement factor become larger. At the same time the smallest value of $\mu_1$ approaches zero and the largest value of $\mu_2$ approaches infinity. If the concentration degree is not too high (e.g., *M*=0.35), we can construct the device through the combination of diamagnetic materials and paramagnetic materials without super-conductors, which means that we can avoid refrigeration, quenching and other drawbacks in super-conductor devices [6-10]. If we want to achieve a very large concentration degree (e.g., *M*=0.01), we can use the combination of super-conductor and ferromagnetic materials [14, 16] to construct it. In this case we can achieve a much higher enhancement, compared with traditional super-conductor based magnetic lenses [6-10].

As the role of our device is to compress the magnetic field, we can simply add another identical device behind the former one to further concentrate the magnetic field (see Fig. 2). The comparison between Fig. 1 and Fig. 2 reveals that two adjacent identical devices can achieve a better magnetic field enhancement at the final output of the structure than a single device, though the improvement is slight. The reason for this will be discussed in section 3. If we want to further increase the degree of the magnetic field enhancement, we will use some other structures which will also be given in section 3.

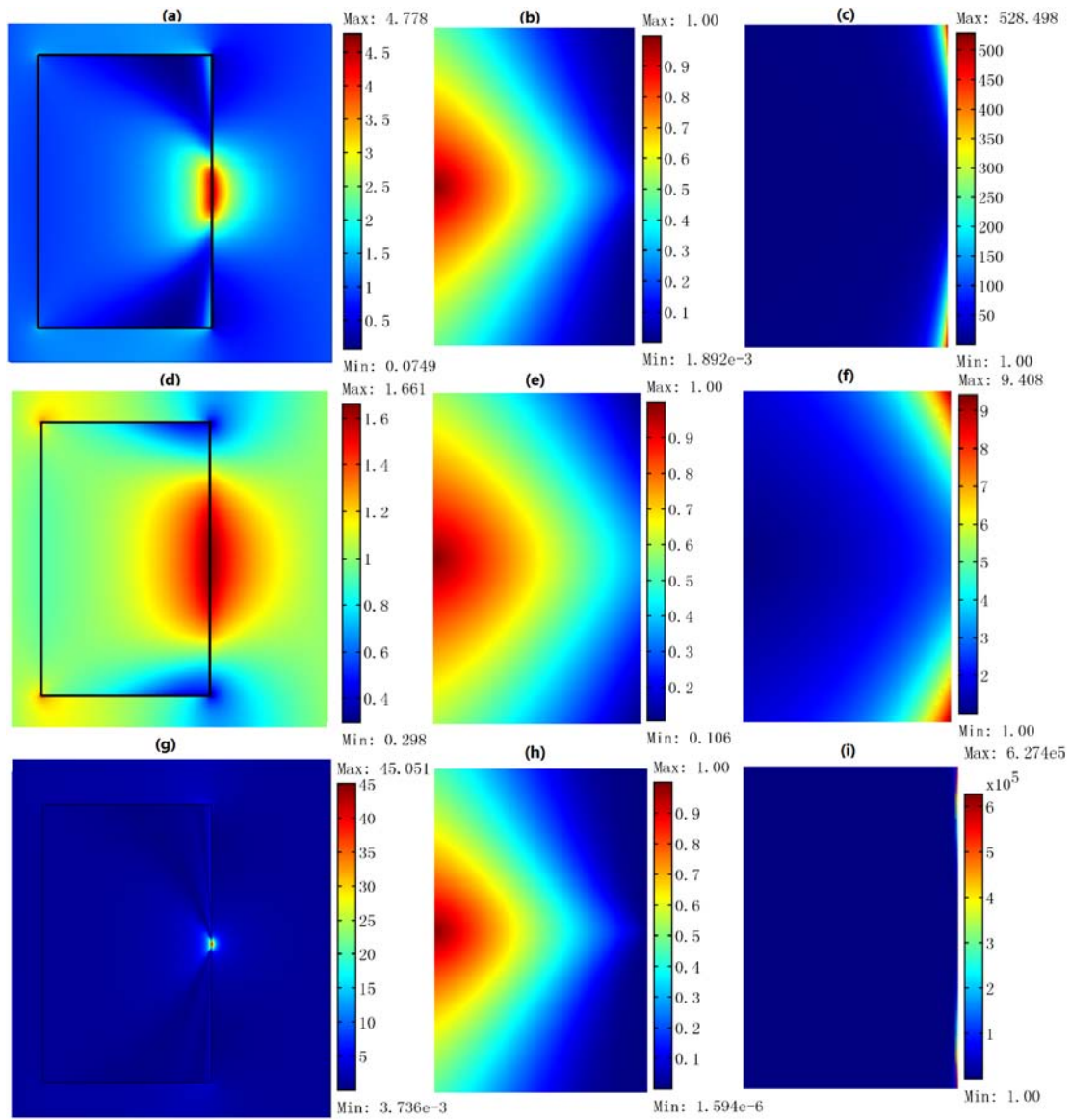

Fig. 1.2D FEM simulation results. The height of the device is 0.8m and the thickness of the device is $d$=0.5m. When a uniformed background static magnetic field $B_b$=1T is imposed onto the device from left to right, the distribution of the total magnetic flux density is shown in (a) for $M$=0.1, (d) for $M$=0.35 and (g) for $M$=0.01. The principal values of the permeability tensor distribution ($\mu_1$ and $\mu_2$) are shown for $M$=0.1 in (b) and (c); for $M$=0.35 in (e) and (f); for $M$=0.01 in (h) and (i). The black rectangle is the designed compressor and the other regions are free space.

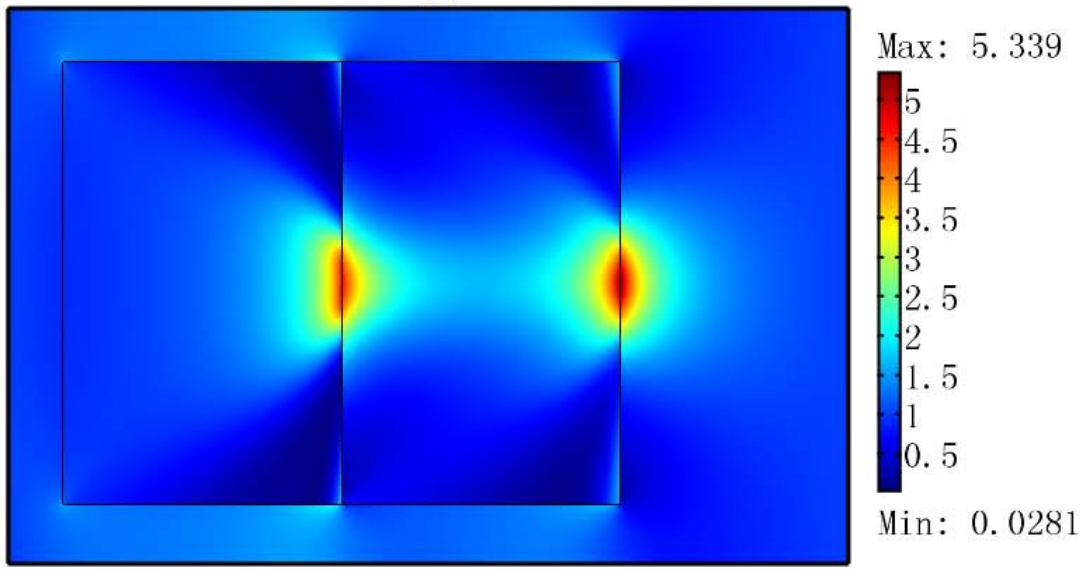
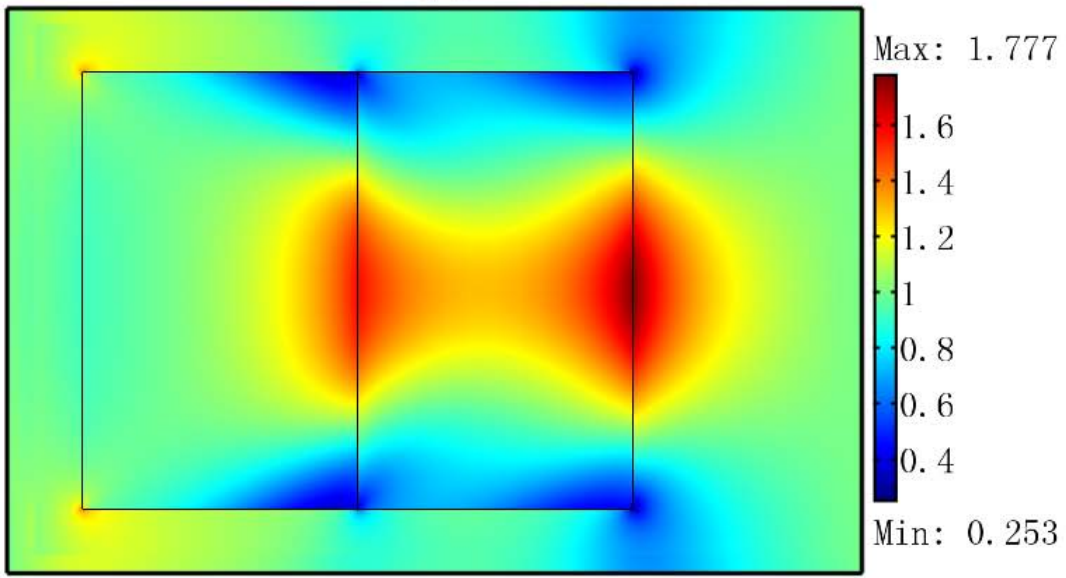
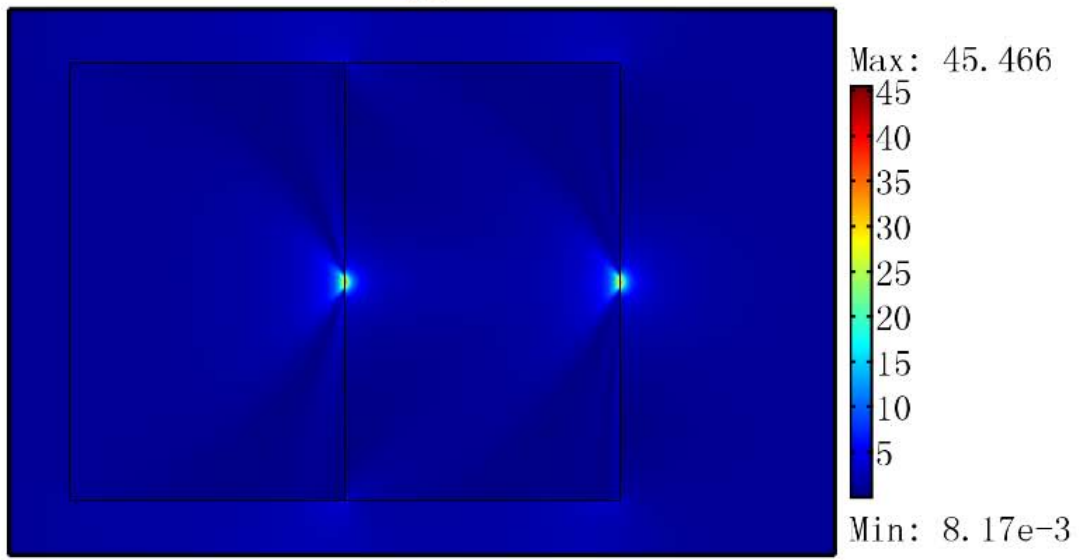

Fig. 2.2D FEM simulation results. When a uniformed background static magnetic field $B_b$=1T is imposed onto the device from left to right, the distribution of the total magnetic flux density (B) for two adjacent identical compressors with the same size of Fig. 1 are shown in (a) for $M$=0.1, (b) for $M$=0.35 and (c) for $M$=0.01. The enhancement factor in this case is about 5.3, 1.7 and 45.4 for (a), (b) and (c), respectively. The two black rectangles are identical compressors and other regions are free space.

## 3. Combination structures for static magnetic field enhancement

We have shown that a single compressor based on FET can concentrate the static magnetic field. In this section, we will combine the compressor with some other structures to achieve a further magnetic field enhancement. The reason why we can only achieve a little improvement in magnetic field enhancement by using two adjacent identical compressors is that no matter how many compressors we use, once the DC magnetic field exits from the output surface of the last device, the DC field will diverge very quickly in free space. If we can add some other structure which can prevent the DC magnetic field from diverging too quickly, we can achieve a better DC magnetic field enhancement. As diamagnetic materials ($0<\mu<1$) can exclude the DC magnetic field, we can set up some diamagnetic structures after the designed device to prevent the DC magnetic field from diverging once it gets out of the compressor. As shown in Fig. 3(a), we add two parallel slabs with $\mu$=0.3 after the compressor (which is the same as the one in Fig. 1(a)). In this case, we get a better enhancement (the enhancement factor is about 6 in the free space region). For comparison, we also simulate the case that a single structure with two parallel diamagnetic slabs ($\mu$=0.3) is used for static magnetic field enhancement (see Fig. 3(b)). In this case the enhancement factor is only about 1.9 in the free space region.

A magnetic hose which can transport the static magnetic energy to a long distance has been proposed very recently based on TO [24]. A structure with a super-conductor shell whose permeability approaches zero) and an inner ferromagnetic core (whose permeability approaches infinity) has been shown to be able to guide the static magnetic energy to an arbitrary distance. We can combine our concentrator with the reduced magnetic hose to obtain a further magnetic field enhancement. In order to avoid using superconductors, as we have mentioned its drawbacks in the introduction, we can reduce the magnetic hose by using a diamagnetic shell whose permeability is smaller than 1 but not necessarily very close to zero and a paramagnetic core whose permeability is larger than 1 but not necessarily infinitely large. Then we can put our compressor (the same as the one in Fig. 1(a)) before the reduced magnetic hose and remove some paramagnetic materials in the middle region of the reduced hose to get a higher magnetic field enhancement in this empty/free region (the enhancement factor is about 8.5 in this free space region; see Fig. 3(d)). For comparison, we also simulate the case when a single reduced magnetic hose structure is used for static magnetic field enhancement (see Fig. 3(c)). In this case the enhancement factor is only about 3.5 in the middle free space region. These combination structures can be utilized to enhance the sensitivity of magnetic sensors.

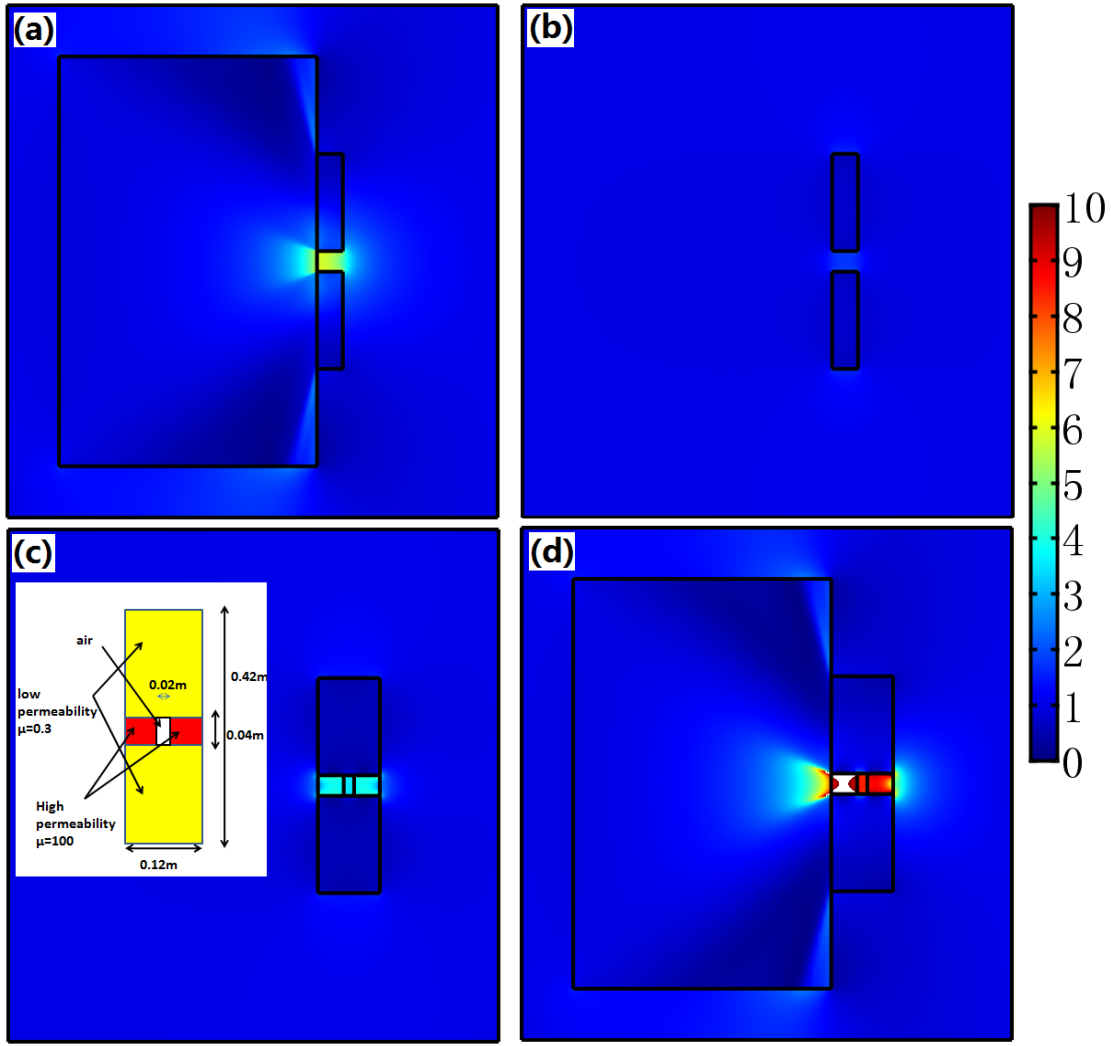

Fig. 3.2D FEM simulation results. When a uniformed background static magnetic field $B_b$=1T is imposed onto the whole structure from left to right, the distribution of the total magnetic flux density is shown in (a) when the structure is the combination of the compressor (with the same size as Fig.1(a)) and two parallel slabs with $\mu$=0.3. The height and thickness of each slab is 19cm and 5cm respectively. (b) when the structure is only two parallel slabs with the same size and permeability as Fig. 3(a). (c) when the structure is only the reduced magnetic hose. (d) when the structure is the combination of the compressor and the reduced magnetic hose. The reduced magnetic hose is shown in detail in the inset of Fig. 3(c). The white region means that the magnetic field is larger than the largest value of the color bar.

4. **Discussion and summary**

Based on FET, we have designed a novel compressor for static magnetic field enhancement. A simulation based on FEM shows good performance of our device. Compared with other devices based on TO for static magnetic field concentration [16, 17], our device is not a closed structure, which has potential applications in some special cases. For example, magnetic nano-particles, attached by cytotoxic drugs for targeted chemotherapy or therapeutic DNA to correct a genetic defect, have been utilized *in vitro* and in animal experiments in recent years [3, 4]. A magnetic field with high strength and high gradient is needed in this case to allow magnetic nano-particles to be captured and extravasated at the target. While this may be effective for targets close to the body's surface, as the magnetic field strength falls off rapidly with distance, it is very difficult to send them to deeper targets within the body. A stronger magnetic field with high gradient can break through this barrier and acquire better control of magnetic nano-particles inside the human body. Our magnetic compressor is an open structure and can also achieve high magnetic fields with high gradient, which will have potential applications for controlling magnetic nano-particles for gene and drag delivery.

Our compressor can also be combined with other structures to get an even higher magnetic field enhancement. More importantly, the permeability values for all the materials used in our compressor are all larger than 0, which means it can be realized by currently available materials or DC meta-materials. We should also note that some meta-materials without magnetic materials can also achieve a magnetic response even for DC magnetic fields. Those meta-materials, combined with high permeability materials (e.g., soft ferromagnetic materials [14, 16]) can be utilized to construct our device.